\documentclass[10pt]{iopart}
\usepackage{epsfig,graphics,color}

\newcommand \ie {{\it i.e.} }
\newcommand \f {\not\!}

\newcommand \kd  {\delta}
\newcommand \ra  {\rightarrow}

\newcommand \fp {{\bf p}}

\newcommand \fn {{\bf n}}

\newcommand \fx {{\bf x}}

\newcommand \si {\sigma}

\newcommand \x {\cdot}

\newcommand \A {\alpha}

\newcommand \lc {\langle}
\newcommand \rc {\rangle}
\newcommand \prt {\partial}

\newcommand \nt {\noindent}

\newcommand \al {\alpha}

\newcommand \bvec{\left( \begin{array}{c} }
\newcommand \evec{\end{array} \right)}
\newcommand \trc {\mbox{{\bf Tr}}}
  
\newcommand \bea{\begin{eqnarray} }
\newcommand \eea{\end{eqnarray} }
\newcommand \nn {\nonumber}
\newcommand {\be} {\begin{equation}}
\newcommand {\ee} {\end{equation}}
\newcommand {\epem} {$e^+ e^-$}

\newcommand {\gev} {\mbox{GeV}}

\begin{document}

\article[Dihadron fragmentation function]{Quark Matter 2004}
{Dihadron fragmentation functions and high Pt hadron-hadron correlations} 
\author{A. Majumder} 
\address{Nuclear Science Division, 
Lawrence Berkeley National Laboratory\\
1 Cyclotron road, Berkeley, CA 94720}
\begin{abstract} 
We propose the formulation of a dihadron fragmentation function 
in terms of parton matrix 
elements. 
Under the collinear factorization approximation and facilitated by the
cut-vertex technique, the two hadron inclusive
cross section at leading order (LO) in \epem annihilation 
is shown to factorize
into a short distance parton cross section and the long distance
dihadron fragmentation function. We also derive 
the DGLAP evolution equation of this function at leading log. 
The evolution equation for the 
non-singlet quark fragmentation function is solved numerically
with a simple ansatz for the initial condition and results are
presented for cases of physical interest. 
\end{abstract}



\section{Introduction}


One of the most promising signatures for the formation of a
quark gluon plasma (QGP) in a heavy-ion collision 
has been that of jet quenching \cite{quenching}. This phenomenon 
leads to the suppression 
of high $p_T$ particles emanating from such collisions. Such jet 
quenching phenomena have been among the most striking experimental 
discoveries from the Relativistic Heavy Ion Collider (RHIC) at 
Brookhaven National Laboratory. 

In the investigation of jet suppression, correlations between two high $p_T$ 
hadrons in azimuthal angle are used to study the change of jet 
structure \cite{adl03}. 
While the back-to-back correlations are suppressed in central $Au+Au$ 
collisions, indicating parton energy loss, the same-side 
correlations remain approximately the same as in $p+p$ and $d+Au$ collisions.
Given the experimental kinematics, this is considered as an indication
of parton hadronization outside the medium. However, since the
same-side correlation corresponds to two-hadron distribution
within a single jet, the observed phenomenon is highly nontrivial.
To answer the question as to why 
a parton with a reduced energy would give the same two-hadron
distribution, one has to take a closer
look at the single and double hadron fragmentation functions and
their modification in medium. In this paper, we take the first
step by studying the dihadron fragmentation function in the
process of $e^+e^-$ annihilation.

For reactions at an energy scale much above $\Lambda_{QCD}$, 
one can factorize the cross section into a short-distance 
parton cross section which is computable order by order 
as a series in $\al_s(Q^2)$ and a long-distance phenomenological 
object (the fragmentation function)
which contains the non-perturbative information of parton 
hadronization~\cite{col89}. 
These functions are process independent. 
If measured at one energy scale, they may be predicted for all other 
energy scales via  the Dokshitzer-Gribov-Lipatov-Altarelli-Parisi (DGLAP) 
evolution equations \cite{gri72}.
In this article, we will extend this formalism to the 
double inclusive fragmentation function $D_q^{h_1,h_2}(z_1,z_2,Q^2)$
or the dihadron fragmentation function, where $z_1,z_2$ represent 
the forward momentum fractions of the two hadrons.


\section{The double fragmentation function and its evolution }


In this article, we will focus on the following semi-inclusive process 
\bea
e^+ + e^- \ra \gamma^* \ra h_1 + h_2 + X \nonumber
\eea
of $e^+ e^-$ annihilation.
We consider two-jet events where both the 
identified hadrons $h_1$ and $h_2$ emanate from the same jet. 
At leading order in the strong coupling, this occurs 
from the conversion of the virtual photon into a 
back-to-back quark and anti-quark pair which 
fragment into two jets of hadrons. 

In the limit of very large $Q^2$ of the reaction, we may 
invoke the collinear approximation. Under this approximation, 
at leading twist,  
we can demonstrate the factorization of 
the two-hadron inclusive cross
section into a hard total partonic cross section $\si_0$ and 
the double inclusive fragmentation function $D^{h_1 h_2}_{q}(z_1,z_2)$ (see 
Ref. \cite{maj04} for details),

\bea
\frac{d^2 \si}{dz_1 dz_2} = \sum_{q} \si_0^{q\bar{q}} 
\left[ D_{q}^{h_1 h_2} (z_1,z_2) + 
D_{\bar{q}}^{h_1 h_2} (z_1,z_2) 
\right].
\label{LO_Dz1z2}
\eea

\nt In the above equation, the leading order 
double inclusive fragmentation function of a quark is obtained as

\bea
\!\!\!\!\!\!\!\!\!\!\!\!\!\!D_q^{h_1,h_2}(z_1,z_2) &=& \int 
\frac{dq_\perp^2}{8(2\pi)^2}  \frac{z^4}{4z_1z_2}
\int \frac{d^4 p}{(2\pi)^4}   
\trc \bigg[ \frac{ \f n}{2 \fn \x \fp_h} 
\int d^4 x e^{i\fp \x \fx} \sum_{S - 2} \nn \\
& & 
\lc 0 | \psi_q^\alpha (x) | p_1 p_2 S-2 \rc  
\lc p_1 p_2 S-2 | \bar{\psi}_q^\beta (0) | 0 \rc \bigg]
\kd \left( z - \frac{p_h^+}{p^+}  \right) .\label{dihad_def} 
\eea

\begin{figure}[htb!]
\hspace{0cm}
  \resizebox{2.4in}{2.4in}{\includegraphics[0in,0in][6.0in,6in]{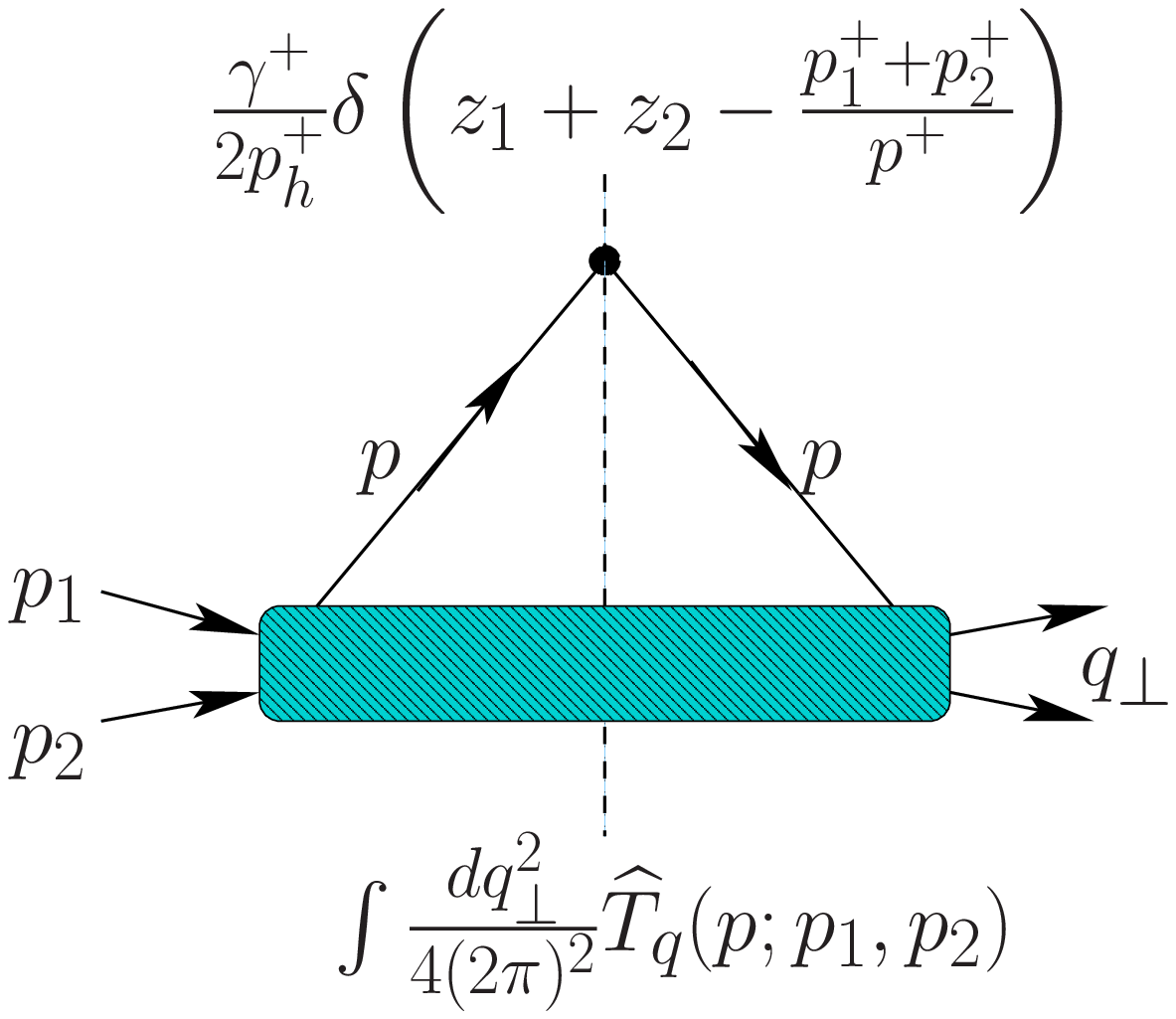}} 
\hspace{0.25cm}
 \resizebox{2.15in}{3.15in}{\includegraphics[0.5in,1in][5.5in,9in]{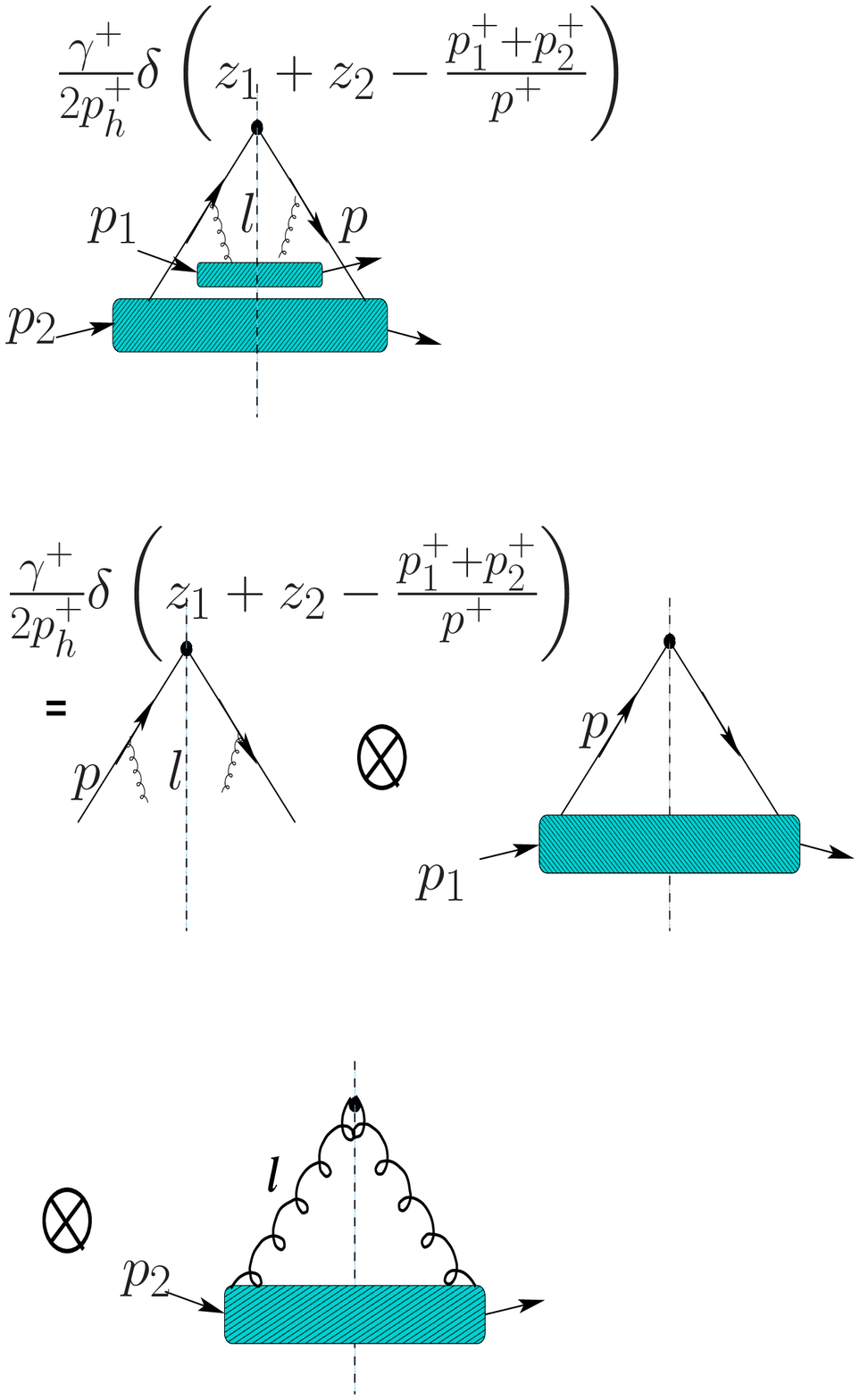}} 
  \caption{The left panel represents the 
  cut-vertex representation of the dihadron fragmentation
    function. The right panel represents a Next-to-Leading order 
    correction from quark and gluon single fragmentation.}
    \label{cutvert1}
\end{figure}

\nt
In the above equation $z=z_1+z_2$, $\fp$ represents the momentum of the 
fragmenting parton, $\fp_h$ is the sum of the momenta of the two detected hadrons 
\ie $\fp_1 + \fp_2$ and $\fn$ is a light-like null vector. The sum over 
$S$ indicates a sum over all possible final hadronic states.
The above equation 
may be represented by the diagrams of the cut vertex notation as 
that in the left panel of Fig.~\ref{cutvert1}. Note that all transverse 
momentum $q_\perp$ up to a scale $\mu_\perp$ 
have been integrated over into the definition of the 
fragmentation function.  Hadrons with transverse momenta $\geq \mu_\perp$
may not emanate from the fragmentation of a single parton.

In the interest of simplicity 
we specialize to the case of the non-singlet fragmentation function, 
$D_{NS}^{h_1,h_2} (z_1,z_2) =  D_q^{h_1,h_2} (z_1,z_2) -  
D_{\bar{q}}^{h_1,h_2} (z_1,z_2)$.
With the definition of the dihadron fragmentation functions 
in the operator formalism,
which is shown to factorize from the hard parton cross section at LO, 
we may now evaluate its DGLAP evolution 
by computing the double inclusive cross 
section at next to leading order (NLO).
This constitutes a rather involved procedure; the reader is referred to 
Ref. \cite{maj04} for details. The resulting DGLAP evolution consists of 
two parts. There remains the usual evolution 
due to the radiation of soft gluons and both detected 
hadrons emanating with a small 
transverse momentum off the same parton. Hadron pairs with 
transverse momentum larger than $\mu_\perp$ 
are produced perturbatively, by the independent 
single fragmentation of a quark and a gluon which emanate from the 
splitting of a quark. The contribution of this process to the evolution
of the dihadron fragmentation function is represented by the cut-vertex 
diagram in the right panel of Fig. \ref{cutvert1}. Incorporating this 
diagram, we obtain the DGLAP evolution of the dihadron fragmentation 
function as 

\bea 
\fl \frac{\prt D_{NS}^{h_1 h_2} (Q^2)}{\prt \log{Q^2}} = 
\frac{\A_s}{2\pi} \Bigg[ P_{q\ra q g} * D_{NS}^{h_1 h_2} (Q^2)  
+ \hat{P}_{q \ra q g} 
\bar{*} D_{NS}^{h_1} (Q^2) D_g^{h_2} (Q^2)  + 1 \ra 2 \Bigg], \label{ns_dglap}
\eea

\begin{figure}[htb!]
\hspace{0cm}
  \resizebox{2.5in}{2.5in}{\includegraphics[0.5in,1in][7.5in,9.5in]{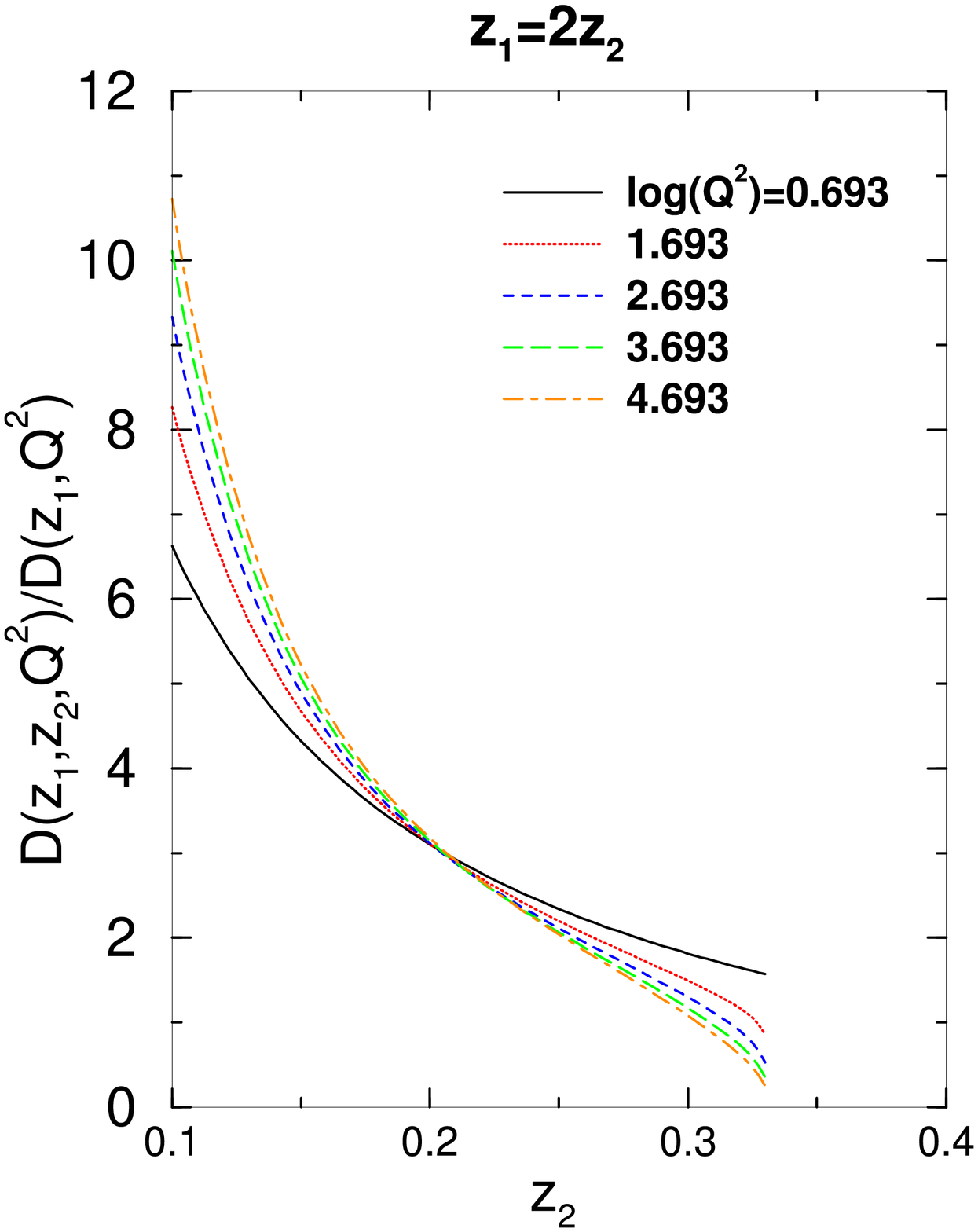}} 
\hspace{0.25cm}
 \resizebox{2.5in}{2.5in}{\includegraphics[0.5in,1in][7.5in,9.5in]{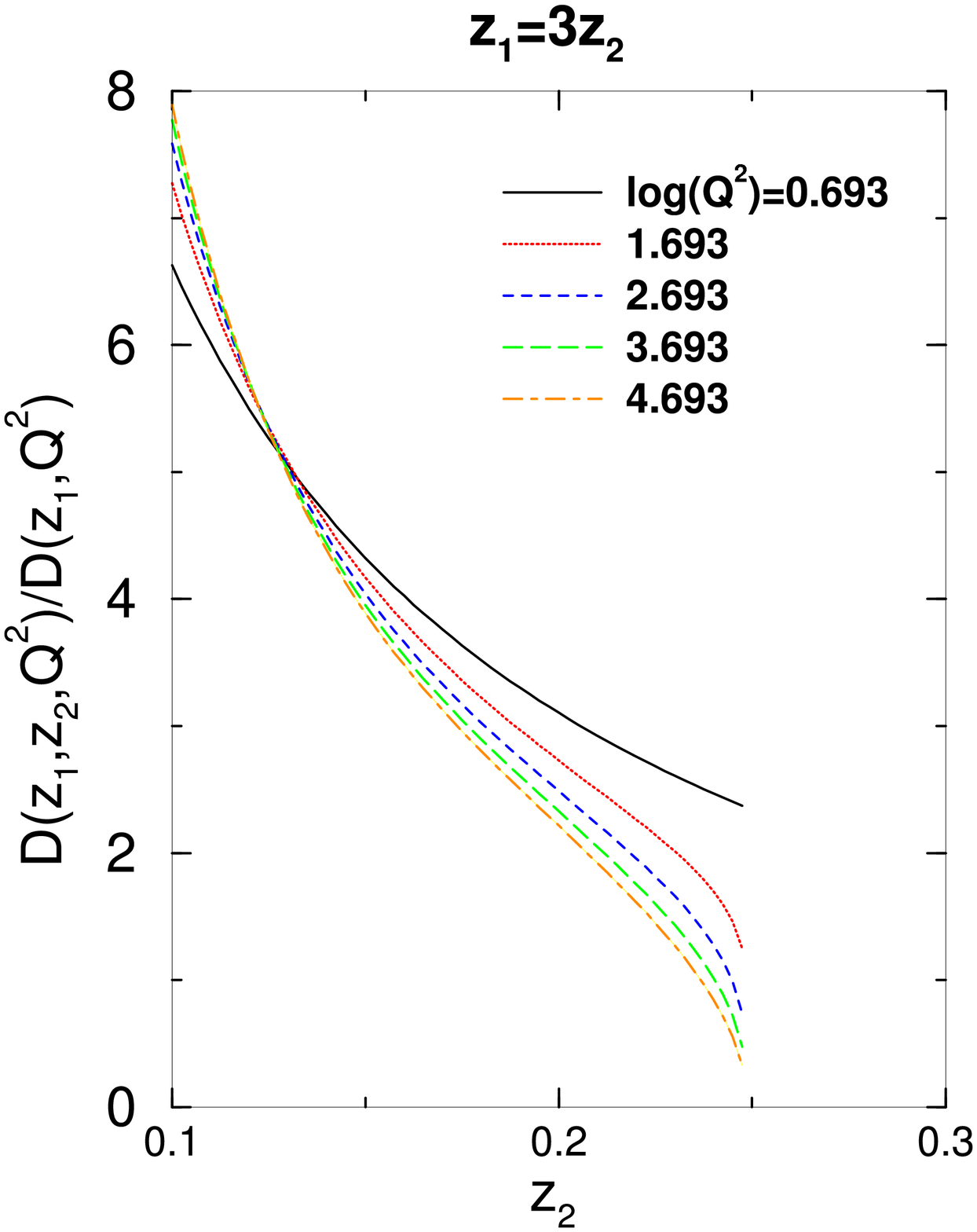}} 
    \caption{Results of the ratio of the non-singlet quark dihadron 
    fragmentation function $D_q^{h_1h_2}(z_1,z_2,Q^2)$ to the single 
    leading fragmentation function  $D_q^{h_1}(z_1,Q^2)$. In the 
    left panel $z_1=2z_2$, in the right panel $z_1=3z_2$. Results are 
    presented from 
    $Q^2= 2 \gev^2 $ to $109 \gev^2$. }
    \label{res3}
\end{figure}

\nt
where the switch $1 \ra 2$ is meant solely for the last term. The 
expressions for the splitting functions ( $P_{q\ra qg}$ and  
$\hat{P}_{q\ra qg}$ ), single fragmentation functions, 
as well as the convolution notations ($*$ and $\bar{*}$)
may be obtained from Ref. \cite{maj04}. The above equation may be solved 
numerically and results for the cases of $z_1=2z_2$ and $3z_2$ are presented in 
Fig.~\ref{res3}. We assume a simple ansatz for the initial condition at 
$Q^2 = 2$ GeV$^2$ \ie $D_{NS}^{h_1 h_2}(z_1,z_2) = D^{h_1}(z_1)D^{h_2}(z_2)$. 
We present results for the evolution with $\log(Q^2)$ at intervals of 1, up to 
$\log(Q^2) = 4.693$ \ie $Q^2 = 109$ GeV$^2$.

\section{Discussions and Conclusions}

This study was motivated by the observation \cite{adl03} that the
same side correlations of two high $p_T$ hadrons in central
$Au+Au$ collisions remain approximately unchanged as compared
with that in $p+p$ and $d+Au$ collisions. Neglecting the differences
in production cross section and fragmentation functions for
different parton species, the integrated yield of the same side 
correlation over a small range of angles should be the ratio of the dihadron to 
the single hadron fragmentation functions, 
$D_a^{h_1h_2}(z_1,z_2,Q^2) / D_a^{h_1}(z_1,Q^2)$, with $z_1$ and
$z_2$ being the momentum fractions of the triggered hadron
and associated hadrons, respectively.

Under the assumption that medium modification due 
to multiple scattering and induced gluon radiation closely 
resemble that of radiative corrections due to evolution in 
vacuum \cite{guowang}, we evaluated the influence of DGLAP evolution on the 
ratio $D_q^{h_1h_2}(z_1,z_2,Q^2) / D_q^{h_1}(z_1,Q^2)$. 
Our numerical results indeed show little change of the ratio 
as $Q^2$ is varied over a wide range of values.
The evolution is shown to be strongly dependent, however, 
on the ratio of the momentum fractions of the two hadrons.
In the results of Ref.~\cite{adl03} the ratio $r=z_1/z_2$ is 
essentially integrated over all values $\geq 1$, as a 
result the effects of the evolution are washed out. Results 
of the evolution of the more complicated singlet fragmentation 
functions as well as the influence of medium modifications will be
presented in a future effort.

\end{document}